\documentclass[11pt]{aime}
\usepackage{booktabs}
\usepackage{array}
\usepackage{xcolor}

\Crefname{appendix}{Appendix}{Appendices}
\crefname{appendix}{appendix}{appendices}

\usepackage{amsmath,amssymb}
\usepackage{booktabs}
\usepackage{array}
\usepackage{graphicx}
\usepackage{xcolor}
\usepackage{microtype}
\usepackage{hyperref}
\usepackage{url}
\usepackage{caption}
\usepackage{natbib}

\definecolor{todored}{RGB}{170,20,35}

\newcommand{\method}{\textsc{RetroTAC}}

\title{RetroTAC: Learning Route-Derived Synthetic Accessibility for PROTACs}

\correspondence{\email{rocio.mercado@chalmers.se}}
\codeavailability{\url{https://github.com/ribesstefano/RetroTAC}}
\date{\today}

\author[1]{Stefano Ribes}
\author[1]{Jia Xin Zhu Zhu}
\author[1]{Tingting Mo}
\author[1]{Lukáš Hudec}
\author[2]{Kostas Papadopoulos}
\author[3]{Atanas Patronov}
\author[1,4,*]{Rocío Mercado}

\affiliation[1]{AI Laboratory for Molecular Engineering (AIME), Department of Computer Science and Engineering, Chalmers University of Technology \& University of Gothenburg, Gothenburg, Sweden}
\affiliation[2]{Generative Matter, Bulgaria}
\affiliation[3]{Independent researcher}
\affiliation[4]{Science for Life Laboratory (SciLifeLab), Gothenburg, Sweden}

\contribution[*]{Corresponding author}

\abstract{
Proteolysis-targeting chimeras (PROTACs) enable targeted protein degradation, expanding therapeutic opportunities beyond conventional inhibition. However, assembling their large, multifunctional structures creates synthetic challenges that hinder their discovery. Traditional synthetic-accessibility scores can overestimate this difficulty by penalizing PROTACs' molecular complexity while overlooking modular assembly from accessible building blocks. In contrast, retrosynthesis planning tools provide explicit routes but are computationally expensive, particularly for large molecules. Here, we introduce a human-interpretable, PROTAC-oriented route-derived score that rewards short synthetic depth and balanced joining of molecular fragments, and \method{}, a machine-learning surrogate that predicts this score directly from molecular structures. \method{} combines complementary models in an ensemble, achieving a root mean squared error of 0.13 and an $R^2$ of 0.65 on held-out molecules. We further evaluate its utility on PROTACs enumerated by recombining warheads, linkers, and E3-ligase ligands, followed by retrosynthetic planning and route analysis. Candidates selected for high predicted scores achieve a mean route-derived synthesizability of 0.73, compared with 0.31 for low-score candidates; 98.75\% versus 51.00\% are fully resolved under the tested search settings. These results support \method{} as a fast prioritization tool for virtual PROTAC libraries, connecting molecular screening with interpretable route characteristics while reserving explicit planning for selected candidates.
}

\begin{document}
\maketitle

\section{Introduction}
\label{sec:introduction}

Proteolysis-targeting chimeras (PROTACs) are heterobifunctional molecules that recruit a protein of interest to an E3 ubiquitin ligase, inducing target degradation through the ubiquitin--proteasome system \citep{hu_recent_2022,gharbi_comprehensive_2024}. A PROTAC contains a target-binding warhead and an E3-ligase ligand connected by a linker. This modular architecture offers a route to modulating proteins that are difficult to drug with occupancy-driven inhibitors, but it also produces molecules that are larger and more flexible than conventional small-molecule drugs \citep{weerakoon_impact_2022,price_beyond_2024}. Synthesis is therefore a practical bottleneck, especially as generative models can enumerate virtual degraders much faster than routes can be assessed or executed \citep{nori_novo_2022,li_3d_2023,guo_link-invent_2023,xu_design_2026}.

Fast synthetic-accessibility scores are attractive filters, but they prioritize different properties of synthesis routes. Fragment-frequency heuristics assess whether a structure resembles common chemistry; reaction-corpus models infer relative complexity or feasibility; and retrosynthesis-derived classifiers estimate whether one planner succeeds under fixed settings \citep{ertl_estimation_2009,vorsilak_syba_2020,chen_estimating_2024,coley_scscore_2018,neeser_fsscore_2024,thakkar_retrosynthetic_2021}. Most were developed primarily in small-molecule-drug-like chemical space, so their size penalties, fragment vocabularies and reaction corpora may mischaracterize large molecules assembled by joining a small number of modular fragments, as in a PROTAC. Two efforts target PROTACs more directly: DeepPSA is trained specifically on PROTAC synthetic-accessibility data \citep{zhang_deeppsa_2025}, and FSscore has been fine-tuned on a small set of expert-labeled PROTAC pairs \citep{neeser_fsscore_2024}.

Retrosynthetic planning tools provide richer evidence---including step count, convergence, proposed transformations and terminal building blocks---but cost substantially more than a single property prediction and cannot be easily integrated in generative pipeline or virtual screening campaigns.
Their cost is even greater when processing large, complex molecules like PROTACs. 
Moreover, found retrosynthesis routes are conditional on the planner, reaction templates, stock, search budget and stochastic seed \citep{chen_retro_2020,thakkar_datasets_2020,maziarz_re-evaluating_2025}.
We therefore reframe route assessment as single-molecule property prediction. We estimate synthesizability directly from a target PROTAC and without calling the planner at inference, using a transparent heuristic designed to summarize a route from a fixed external pipeline.

Our contributions to address the above limitations are threefold:

\begin{itemize}
    \item we define a human-interpretable and PROTAC-oriented route-derived score that rewards short synthesis depth and balanced joining of reactants;
    \item we develop and evaluate a fast and accurate proxy model for retrosynthesis called \method{};
    \item and we openly release nearly 20,000 PROTAC synthesis routes generated by ShallowTree, as well as the \method{} ensemble model.
\end{itemize}

\section{Related work}
\label{sec:related}

Two prior scores address PROTACs directly. FSscore fine-tunes a general feasibility ranker on pairwise expert preferences over PROTAC-DB \citep{neeser_fsscore_2024,weng_protac-db_2023}; tuning shifts its scores toward the easy end and narrows the gap between a PROTAC and its most complex fragment, consistent with joining the three modules not being the hard step, but it draws on only tens of expert-labeled pairs. DeepPSA \citep{zhang_deeppsa_2025}, the only model so far built for PROTAC synthetic accessibility, is a graph classifier that decides easy versus hard over a proprietary set of 3,644 PROTACs. Its authors release trained weights, so we can run it on our own molecules (Appendix~\ref{app:deeppsa}), though its accuracy does not compare directly with a route-derived score.
 
Route scores draw on information molecule-only models lack, such as graph-edit distance with a length penalty \citep{gao_retroscore_2025}, template $n$-gram similarity to known routes \citep{li_retro-bleu_2024}, step penalties with forward-model confidence \citep{kreutter_multistep_2023}, and recursive precursor cost with expert grades \citep{sun_chemirise_2021}. Ours is the most lightweight of these route scores, and, to our knowledge, the first defined on PROTAC syntheses. It reads only route topology and heavy-atom counts, needs no second reaction model, and---unlike a length-only penalty---separates routes that decorate one large intermediate from routes that join comparably sized modules. While it doesn't capture feasibility, selectivity or yield, it does measure route depth and modular balance.

\section{Methods}
\label{sec:methods}

\begin{figure}[t]
\centering
    \includegraphics[width=1\linewidth]{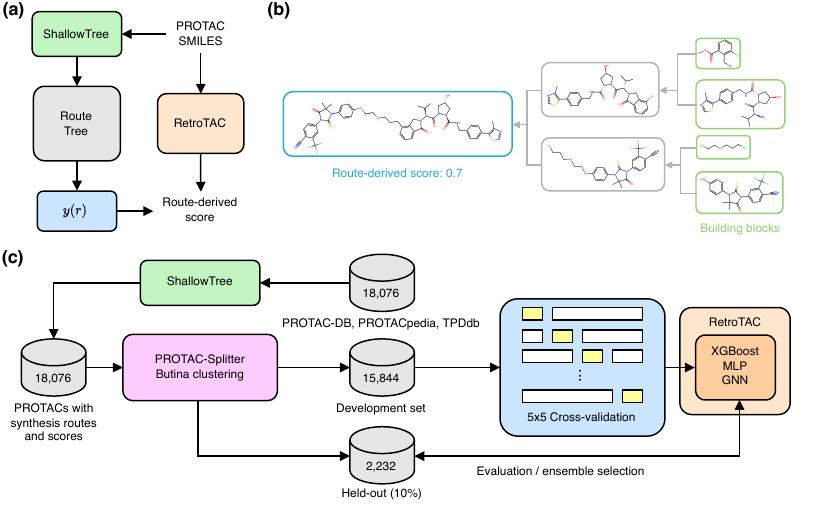}
    \caption{(a) \method{} is trained to proxy the route-derived score computed from synthesis routes found by ShallowTree. (b) Example synthesis route found by ShallowTree. (c) Training and evaluation pipeline. After obtaining the synthesis routes, we isolate a held-out set via Butina clustering. The remaining development set is used to train and compare three surrogate models in a repeated cross-validation setting.}
\label{fig:overview}
\end{figure}

We frame the learning objective as a regression task.
Each sample contains a target PROTAC SMILES $x_i$ and a route $r_i$ supplied by an external retrosynthesis tool. The route determines the training label, which we refer to throughout as the route-derived score, but the predictor receives only the target molecule and learns $f_\theta(x_i)\approx y(r_i)$ (Figure \ref{fig:overview}a). We use this to create a surrogate for route planning: explicit routes are required to create training data, but not to score a new molecule.

\subsection{Data curation}

We collected and standardized 21,189 PROTAC SMILES from PROTAC-DB~\citep{ge_protac-db_2025}, PROTACpedia~\citep{london_protacpedia_2024}, and TPDdb~\citep{qin_tpddb_2026}. We generated synthesis routes for these using ShallowTree, resulting in 18,076 PROTACs with route-derived labels. We used 15,844 molecules for model development and set aside 2,232 ($\sim$10\%) for a final held-out evaluation. The held-out set was isolated before model fitting with adaptive Butina clustering~\citep{butina_unsupervised_1999} on 2,048 Morgan fingerprints, radius 8, at a Tanimoto-distance \citep{lipkus_proof_1999} cutoff of 0.85 (Appendix~\ref{app:heldout_clustering}). Figure~\ref{fig:overview}c illustrates the \method{} training and evaluation pipeline. 
The development set (i.e., train and validation sets) requires a second safeguard against data leakage before cross-validation. PROTACs are assembled from recurring warheads, linkers and E3 ligands; we therefore decomposed every molecule with PROTAC-Splitter~\citep{ribes_protac-splitter_2026} and grouped the development folds by Bemis--Murcko scaffolds~\citep{bemis_properties_1996} of the warhead (Appendix~\ref{appendix:scaffold}).
For constructing the Caruana ensemble (Section \ref{sec:methods:ensemble}) and comparing the other strategies, we isolated a hill-climb set of 463 (20\%) randomly sampled data from the held-out set, and evaluated on the remaining 1,854 data points.

\subsection{Retrosynthesis route prediction}

In order to produce the synthetic routes for each of the PROTACs datapoints we used a proprietary retrosynthesis planning tool ShallowTree. 
ShallowTree is a synthesis planning tool that can generate synthetic routes for a query molecule at a user defined depth. It is inspired by the tools for single-molecule retrosynthesis guided by neural network expansion policy such as AiZynthFinder~\citep{genheden_aizynthfinder_2020}. 
Depending on user's preferences, the search algorithm can balance depth first over breadth first search. ShallowTree expansion and filter policies as well as the building block collections are compatible with the ones used by AiZynthFinder. However, ShallowTree is designed for the purpose of processing multiple target molecules in parallel while also keeping a minimal memory and optimal CPU usage. This makes it particularly suitable for processing rather challenging targets such as PROTACs even when the dataset consists of tens of thousands of molecules. All searches used the eMolecules building-block collection as purchasable stock~\citep{emolecules_emolecules_nodate}. We processed the entire dataset data set and successfully resolved about 92.4\% or 17,802 out of 19,274 PROTACs molecules by using 64 CPUs with a timeout limit of 30 min per molecule. The maximum permitted depth was 12. The remaining 1,472 unresolved molecules exceeded the allocated timeout or exhausted the available trees at depth 12. Generating the synthetic trees for the entire dataset on 64 CPUs took less than 50h. An example of a synthetic tree generated by ShallowTree can be found on Figure \ref{fig:overview}b, and additional routes in Appendix \ref{app:routes}.

\subsection{Route-derived synthesizability score definition}
\label{subsec:score}

We designed the route-derived score around two key concerns in PROTAC synthesis: \textit{how many sequential reactions are required before all branches reach purchasable building blocks}, and, \textit{when the route joins fragments, does it combine similarly sized modules or iteratively decorate one large intermediate with small groups?} The first property is represented by the longest linear sequence in the route, also known as \textit{route depth}, denoted here as $D(r)$, which decreases linearly to zero at a fixed saturation depth $D_{\mathrm{max}}$. The second is represented by a balance term $B(r)$: at each bimolecular reaction step, we compare the heavy-atom counts of the joined fragments using the ratio of the smaller to the larger fragment, and then average over the route. The final score is
\begin{equation}
y(r)=\frac{1}{2}\left[\max\left(0, 1-\frac{D(r) - 1}{D_{\mathrm{max}} - 1}\right)+B(r)\right],\qquad D_\mathrm{max}=11.
\label{eq:target}
\end{equation}
Both the depth component and $B(r)$ lie in $[0,1]$, where larger values describe shorter, more balanced routes.
This two-term form emerged from a simpler design choice rather than a fitted optimization.
A detailed analysis of the scoring design is reported in Appendix \ref{app:score_analysis}.

Two boundary cases do not have a route to score. Targets already present in the purchasable stock receive $y=1$, whereas searches that fail to reach stock within the planner budget receive $y=0$. These endpoints are not symmetric: one records known availability, while the other may reflect either genuine difficulty or a limitation (e.g., timeout, or max depth reached) of the search.

\subsection{Selected models and statistical comparison}
\label{sec:models}

We explored three model architectures for predicting the route-derived score ($y$): XGBoost (XGB) \citep{chen_xgboost_2016}, multilayer perceptron (MLP), and Chemprop (GNN) \citep{graff2026chemprop}.
Both XGB and MLP receive 1,024-bit, radius-8 Morgan fingerprints with chirality enabled, alongside 2D RDKit descriptors. Fingerprints are eventually reduced with truncated SVD via hyperparameters optimization, while descriptors are median-imputed, constant columns are removed, and the remaining values are standardized; the GNN instead receives the molecular graph parsed directly from SMILES.
See Appendix \ref{app:models} for details.

To compare the three model backbones, we follow the repeated-cross-validation (CV) framework of \citet{ash_practically_2025}, giving 25 paired evaluations per model. For every outer fold, hyperparameters and early stopping are chosen on a separate grouped by warhead scaffolds 80/20 split of the available training molecules; the outer fold remains untouched until final scoring. We then refit each model on the full development set and evaluate it once on the held-out set.

Statistical comparisons of the 25 performance measurements from CV are performed with AutoRank \citep{herbold_autorank_2020} at a significance level of 0.05. To select the testing procedure, we assess between-model differences in fold-score variance using Levene's test and the ratio of the largest to the smallest variance. We flag substantial heteroscedasticity only when Levene's $p<0.05$ and this ratio exceeds 9~\citep{blanca_effect_2018}. Otherwise, we explicitly select AutoRank's parametric procedure: repeated-measures ANOVA followed by Tukey HSD for pairwise comparisons with family-wise error control. This branch assumes approximate normality regardless of the normality diagnostic results. When the variance criterion is exceeded, we retain AutoRank's automatic test selection based on its normality and variance-homogeneity checks.

We evaluate regression performance primarily using $R^2$ and RMSE, complemented by MAE for absolute prediction error and Spearman's $\rho$ for ranking performance. For a secondary classification analysis, we treat route-derived scores of $\ge 0.7$ as positive and apply the same threshold to model predictions. These binary labels are used to compute MCC, F1, precision, recall, specificity, balanced accuracy, accuracy, Cohen's $\kappa$. ROC-AUC and PR-AUC instead use continuous predictions against the binary reference labels.

\subsection{Ensemble selection}
\label{sec:methods:ensemble}

After identifying the best performing models, we compared three strategies for selecting cross-validated models for an ensemble: uniform weighting of all $5\times5$ CV models and architectures (XGB, MLP, and GNN), for a total of 75 models; the best backend architecture among XGB, MLP, or GNN, with 25 models; and a \citet{caruana_ensemble_2004} greedy selection over all the 75 models.
The ensembles return mean predictions and uncertainty, measured as standard deviation of individual models' outputs. We refer to the best ensemble model as \method{} throughout this work.

\subsection{Prospective validation}
\label{sec:methods:prospective}

We scored the unseen, enumerated PROTACs from \citet{ribes_protac-splitter_2026} using the \method{} surrogate, recording each molecule's predicted synthesizability (the Caruana-weighted ensemble mean) and its uncertainty (the unweighted standard deviation across ensemble members). 
We predicted 1,303,154 PROTACs in total, after removing 71 samples overlapping with the development set.
We divided the scored molecules by uncertainty at the median into a \textit{confident} pool and an \textit{uncertain} pool. Within each pool, we isolated the 400 lowest- and 400 highest-scoring molecules by predicted synthesizability, yielding four subsets: confident-low and confident-high, and uncertain-low and uncertain-high. We ran ShallowTree on each subset to obtain a potential synthesis route per molecule, scored these routes as in Section \ref{subsec:score}, and compared to the original \method{} predictions.

\section{Results}
\label{sec:results}

\subsection{Score distributions and correlations}
\label{sec:results:score_analysis}

\begin{figure}[t]
\centering
    \includegraphics[width=1\linewidth]{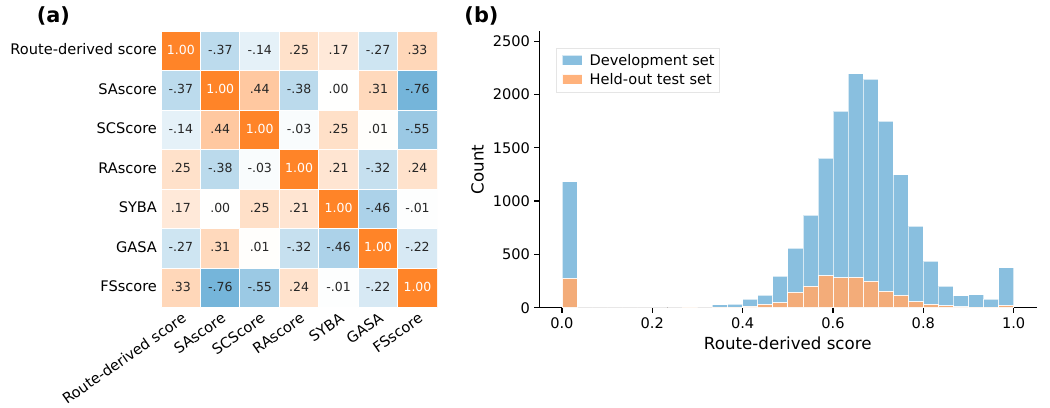}
    \caption{(a) Spearman correlation matrix between our proposed route-derived score and common synthesizability scores developed for small molecules. (b) Development versus held-out set distribution of the route-derived synthesizability score.}
\label{fig:scores}
\end{figure}

In Figure~\ref{fig:scores}a we compare the correlation between our proposed route-derived score and existing synthesizability scores calculated on all the collected PROTACs (SA score, $\rho=-0.37$; SCScore, $\rho=-0.14$; RAscore, $\rho=0.25$; SYBA, $\rho=0.17$; GASA, $\rho=-0.27$; and FSscore, $\rho=0.33$) illustrating weak correlation, and even anti-correlation, with existing metrics, suggesting current metrics which have been mostly designed for small molecule drugs do not transfer well to the PROTAC space. 
We next examine the route-derived synthesizability score developed herein across the development ($n=15,844$) and held-out ($n=2,232$) sets. The held-out distribution was slightly shifted towards lower values, with a larger concentration at zero (Figure~\ref{fig:scores}b); a two-sample Kolmogorov--Smirnov test confirmed the two distributions differ significantly ($D=0.1807$, $p=2.513\times10^{-56}$), likely due to the peaks at 0.

\subsection{Cross-validation and held-out performance}

The GNN achieved the highest mean $R^2$ ($0.509\pm0.056$) and lowest RMSE ($0.142\pm0.010$), followed by MLP ($0.290\pm0.114$ and $0.171\pm0.016$, respectively) and XGB ($0.237\pm0.121$ and $0.177\pm0.018$). Results are summarized in Table~\ref{tab:performance}. The GNN also achieved the lowest mean MAE (0.094) and highest mean Spearman correlation (0.597), with lower variability across folds in $R^2$ and RMSE than either baseline.

\begin{figure}[t]
    \centering
    \includegraphics[width=1\linewidth]{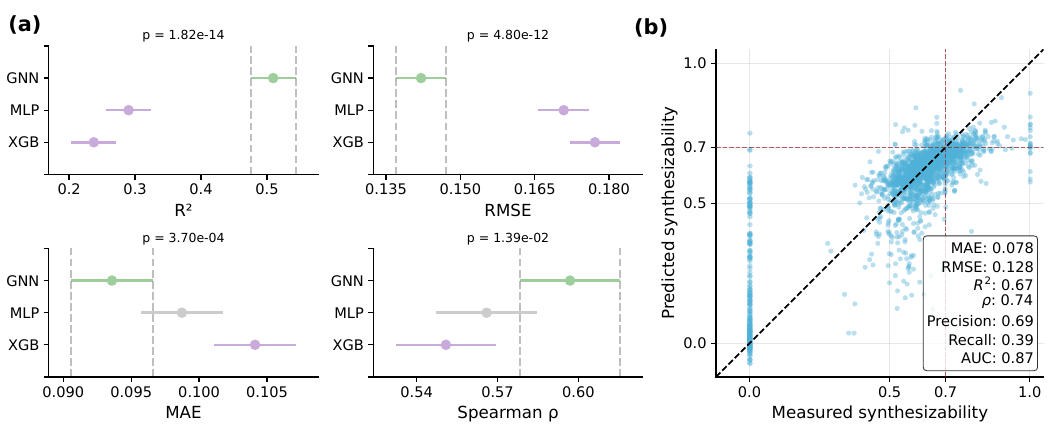}
    \caption{(a) Simultaneous confidence interval plots comparing the three models on regression metrics. (b) Parity plots of the Caruana-based ensemble.}
\label{fig:performance}
\end{figure}

To compare the $R^2$ from different model backbones, Levene's test ($p=0.0877$) and the maximum-to-minimum fold-variance ratio (4.68) met our criterion for selecting the parametric procedure, although the Shapiro-Wilk test rejected normality for MLP ($p=0.001$). Under this procedure, repeated-measures ANOVA detected differences in mean $R^2$ ($p<0.001$), and Tukey HSD identified significant differences between all model pairs at a family-wise significance level of $0.05$, supporting the ordering GNN $>$ MLP $>$ XGB (Figure \ref{fig:performance}a).
We see the same ordering for RMSE, where the GNN is outperforming both MLP and XGB. There is no significant difference between GNN and MLP models on MAE and Spearman correlation instead.
Parametric assumption tests and normality diagnostics are reported in Appendix \ref{app:normality}.

On the 2,232 held-out molecules, the GNN refit on the whole development set retained the highest $R^2$ (0.615) and lowest RMSE (0.140), followed by MLP ($R^2=$0.591, RMSE=0.144) and XGB ($R^2=$0.515, RMSE=0.157).
Ranking performance was similar across models, with Spearman correlations ranging from 0.705 to 0.711. At the classification threshold of 0.7, the GNN achieved ROC-AUC 0.850, precision 0.65, and recall 0.53. XGB had slightly higher precision (0.66) but lower recall (0.20), while MLP achieved precision 0.58 and recall 0.39.
More results are reported in Appendix~\ref{app:remaining-metrics}.

\begin{table}[t]
\centering
\caption{Regression, ranking, and thresholded-classification performance. CV values are mean $\pm$ s.d. over 25 paired outer evaluations. Held-out models are evaluated on 2,232 molecules. ROC-AUC is computed from continuous predictions.}
\label{tab:performance}
\footnotesize
\begin{tabular}{llrrrr}
\toprule
Evaluation & Model & $R^2$ & RMSE & Spearman $\rho$ & ROC-AUC\\
\midrule
5$\times$5 CV & XGB & $0.237\pm0.121$ & $0.177\pm0.018$ & $0.551\pm0.054$ & $0.758\pm0.034$\\
& MLP & $0.290\pm0.114$ & $0.171\pm0.016$ & $0.566\pm0.057$ & $0.764\pm0.033$\\
& GNN & $\mathbf{0.509\pm0.056}$ & $\mathbf{0.142\pm0.010}$ & $\mathbf{0.597\pm0.054}$ & $\mathbf{0.771\pm0.032}$\\
\midrule
Held-out & XGB & 0.515 & 0.157 & 0.705 & 0.840\\
& MLP & 0.591 & 0.144 & \textbf{0.711} & 0.840\\
& GNN & \textbf{0.615} & \textbf{0.140} & 0.709 & \textbf{0.850}\\
\bottomrule
\end{tabular}
\end{table}

\subsection{Ensemble performance}

Ensemble strategies were evaluated on 1,854 molecules, separately from the 463-molecule hill-climb selection subset (Table~\ref{tab:ensemble}). The best single GNN achieved RMSE 0.1378 and $R^2=0.616$. Uniform averaging of all 75 models yielded RMSE 0.1346 and $R^2=0.633$, while averaging the 25 GNNs yielded 0.1332 and 0.641. The Caruana ensemble performed the best, with the reported composition of 13 GNN, seven MLP, and seven XGB members, achieving RMSE 0.1320 and $R^2=0.647$.

The best single model and Caruana ensemble had Spearman correlations of 0.68 and 0.72, respectively; see Figure~\ref{fig:performance}b for parity plots. At the 0.7 threshold, precision was 0.51 and 0.66, while recall was 0.47 and 0.31. Both models assigned positive scores to many zero-labeled molecules, including predictions in the central score range.

Mean member dispersion, a measure of how much ensemble members disagree with each other, was 0.0693 for uniform averaging, 0.0481 for the GNN average, and 0.0675 for Caruana selection; unsurprisingly, the members of the ensemble with the same backbone (GNN) agree most with each other. Spearman correlations between absolute error and member standard deviation were 0.332, 0.369, and 0.337, respectively, suggesting the potential to use ensemble standard deviation as a weakly calibrated error estimate.

\begin{table}[t]
\centering
\caption{Ensemble results on the common 1,854-molecule evaluation subset, disjoint from the 463 cases (20\%) used for Caruana selection (hill-climb set). Dispersion represents the mean disagreement among an ensemble's members, calculated by averaging the individual predictions' standard deviations. The value $\rho(|e|,s)$ is defined as the Spearman correlation between predictions' absolute errors and their corresponding standard deviations.}
\label{tab:ensemble}
\footnotesize
\resizebox{\linewidth}{!}{%
\begin{tabular}{lrlrrrrr}
\toprule
Strategy & $n$ & Composition & RMSE & $R^2$ & Dispersion & $\rho(|e|,s)$ & Cov. $1\sigma/2\sigma/3\sigma$\\
\midrule
Best single & 1 & GNN:1 & 0.1378 & 0.616 & --- & --- & ---\\
Uniform & 75 & XGB:25; MLP:25; GNN:25 & 0.1346 & 0.633 & 0.0693 & 0.332 & 46.9 / 70.3 / 83.7\% \\
Best backend & 25 & GNN:25 & 0.1332 & 0.641 & 0.0481 & \textbf{0.369} & 38.2 / 66.6 / 82.0\% \\
Greedy & 27 & XGB:7; MLP:7; GNN:13 & \textbf{0.1320} & \textbf{0.647} & 0.0675 & 0.337 & 47.0 / 72.9 / 84.4\% \\
\bottomrule
\end{tabular}
}%
\end{table}

\begin{figure}[t]
    \centering
    \includegraphics[width=1\linewidth]{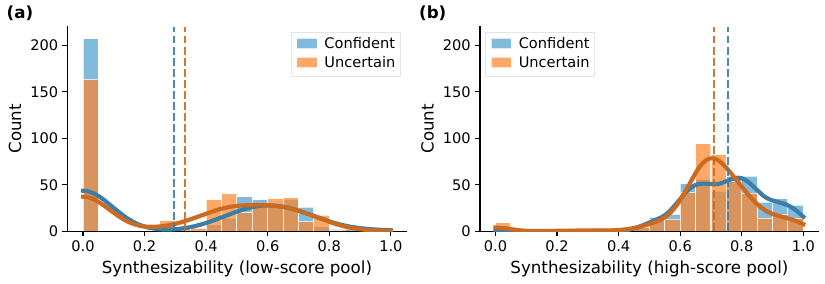}
    \caption{Route-derived synthesizability scores for the four candidate pools ($n$ = 400 each) containing unseen PROTACs, grouped by \method{} ensemble confidence within the low-scored (a) and high-scored (b) tails. Dashed lines mark each subset's mean; curves are kernel density estimates (KDEs).}
    \label{fig:negative_data}
\end{figure}

\subsection{Prospective validation}
\label{sec:prospective}

We evaluated four subsets of 400 unseen PROTACs, stratified by predicted score and certainty. High-score candidates had predictions above 0.868, whereas low-score candidates had predictions below $-$0.058; each pool was divided into confident (high-certainty) and uncertain (low-certainty) subsets.\footnote{\method{} predictions are not clipped to the $[0,1]$ range of the training label, so the lowest-scoring candidates can fall slightly below zero.}

Within the high-score pool (Figure~\ref{fig:negative_data}b), mean route-derived synthesizability was 0.756 for confident candidates and 0.710 for uncertain candidates. Resolution rates were 99.75\% (399/400) and 97.75\% (391/400), respectively, with the latter including 390 fully solved routes and one directly purchasable molecule. Only one confident and nine uncertain candidates yielded no route. Structural summaries of the 399 confident and 390 uncertain non-empty trees showed short routes in both subsets: mean route depths were 2.17 and 2.35, and mean fragment balances were 0.633 and 0.585, respectively.

Within the low-score pool (Figure~\ref{fig:negative_data}a), mean route-derived synthesizability was 0.296 for confident candidates and 0.330 for uncertain candidates. No route was found for 207/400 confident candidates (51.75\%) and 163/400 uncertain candidates (40.75\%), producing a pronounced concentration of scores at zero. Fully solved routes were obtained for 193/400 (48.25\%) and 215/400 (53.75\%), respectively. The uncertain subset also contained 22 partial routes, which were excluded from the resolution rate.
Across the 193 confident and 237 uncertain non-empty trees, mean route depths were 3.45 and 3.94, and mean fragment balances were 0.471 and 0.409, respectively.

\section{Discussion}
\label{sec:discussion}

\subsection{Relationship to existing synthesizability scores}

The modest correlations in Section~\ref{sec:results:score_analysis} indicate that molecule-level scores and our route-derived score capture overlapping but non-equivalent notions of accessibility. Score direction affects the signs but not the small magnitudes, since SA score and FSscore correlate far more strongly with each other ($\rho=-0.76$) than either does with our route-derived score.
 
The reason differs by family. SA score ($\rho=-0.37$) adds an unbounded heavy-atom penalty to its fragment term, so it registers a PROTAC's size rather than how few bimolecular reactions assemble it; the chemist ratings it was fitted to themselves correlate with atom count at $R^2=0.688$. SYBA ($\rho=0.17$) avoids that size bias but still reads only fragment log-odds, with no account of how a molecule is built. SCScore ($\rho=-0.14$) is fitted along single reactions, ordering a product above its reactants but barely separating similarly sized PROTACs of differing route depth. The binary classifiers RAscore ($\rho=0.25$) and GASA ($\rho=-0.27$) return one decision boundary and cannot rank molecules that all lie on its synthesizable side, as we do here. FSscore correlates most strongly ($\rho=0.33$) but encodes a relative preference between molecules rather than a property of a route, and its PROTAC adaptation rests on 100 pairs from a single chemist and is not the default model.
 
Divergence alone does not make the route-derived score the better quantity to rank on; the prospective results do. Among unseen enumerated PROTACs (Section~\ref{sec:prospective}), the candidates \method{} ranked highest were resolved by the planner in 98.75\% of cases against 51.00\% for the lowest-ranked, at a mean route-derived synthesizability of 0.733 against 0.313, and gave shorter, more balanced routes. \method{} therefore offers two advantages the molecule-level scores lack. It discriminates among molecules that are all nominally synthesizable, and a low score is traceable, decomposing into route depth and fragment imbalance. Feasibility, selectivity and yield lie outside what it represents, and we make no claim that it supersedes scores validated against those endpoints.

\subsection{Implications for candidate prioritization}

The ensemble's ranking performance supports its potential use for prioritizing molecules for subsequent route inspection. At the chosen threshold of 0.7, Caruana selection yields a more precise but less comprehensive selection than the best single model. The positive association between member disagreement and absolute error provides an additional signal that could help prioritize predictions for review.

On the enumerated PROTAC candidates, \method{} predicted successful route resolution and routes with fewer steps, shorter sequential depth, and more balanced fragment coupling. These structural differences are consistent with the intended preference of the route score for short, modular syntheses.
The high-low score separation was larger in the confident pool, suggesting that ensemble disagreement may help identify candidates with more reliable extreme predictions. However, the uncertainty distributions overlapped substantially, and selecting score extremes separately within each pool does not isolate the contribution of uncertainty from differences in predicted score or molecular composition.

\subsection{Retrosynthesis planning}

Several properties of the planning tool influence the returned routes used for building the training set of route-derived synthesis scores in this work. Route quality depends primarily on the building-block collection, so the set of returned trees varies substantially across collections, and some collections are better suited to PROTAC synthesis than others; we found the eMolecules collection particularly appropriate here. Notably, however, the expansion policy does not decompose macrocycles, which accounts for a portion of the unresolved PROTACs. Because most molecules were resolved by a search depth of six, we do not expect deeper searches to raise the resolution rate. ShallowTree prioritizes speed and halts once it finds the first complete tree for a molecule, so it retains a single route per target and may overlook better alternatives in exchange for throughput. Furthermore, a minority of routes terminate in building blocks that appear unusual from a synthetic-chemistry standpoint. All of them, however, are cataloged in the eMolecules building-block stock, so we treat them as valid, if initially surprising, starting materials rather than as artifacts of the search. These routes reflect what a retrosynthesis tool, here ShallowTree, proposes, and need not match how a chemist would assemble a PROTAC in practice, which often means preparing the warhead, linker, and E3 ligand separately and joining them through a coupling such as an amide bond formation (for example). We therefore read the scored routes as a first, planner-derived proxy for PROTAC synthesizability rather than as synthesis recipes. A natural extension is to assess the synthesizability of each component separately from that of the assembled PROTAC.

\subsection{Limitations}
\label{sec:limitations}

The route-derived score summarizes a single route from one planner under a fixed stock and budget, and says nothing about experimental feasibility, selectivity, yield, or cost. Its endpoints are asymmetric, since a score of one can simply mean stock membership while a zero conflates real difficulty with a search that failed under the tested settings. The equal weighting of route depth and fragment balance is an uncalibrated design choice, and the 0.7 classification cutoff is a reporting convention, so we treat the continuous score and its ranking as the primary output. Validation against executed syntheses or expert route assessments is needed before reading the score as practical synthetic effort. Further, our search settings match common academic practice and resolved most PROTACs, so we do not attribute the unresolved fraction to under-resourcing. We keep one route per molecule, a standard simplification given that no planner enumerates the route space exhaustively; a resolved route is enough to show a molecule is reachable, which is what the label records. Sampling several routes per molecule, for instance to estimate a success probability, is left to future work. Additionally, the ensemble standard deviation tracks error only weakly and under-covers, so it cannot serve directly as a decision threshold; calibrating it on held-out data would make it more actionable. Finally, we have no experimentally validated negative examples, only enumerated PROTACs of unknown feasibility, so low scores and failed searches do not prove a molecule is hard or impossible to make. Prospective synthesis attempts with recorded failures would provide stronger validation, though structured data of this nature does not exist.

\subsection{Conclusion}

We set out to make a useful synthesizability surrogate for PROTAC assessment, where generative models now propose candidates faster than retrosynthesis models can evaluate them. We defined a transparent route score that rewards short route depth and the balanced joining of fragments, and we trained \method{} to predict it directly from a PROTAC structure, reaching a held-out RMSE of 0.13 and an $R^2$ of 0.65 as an ensemble. On more than one million enumerated PROTACs, the candidates \method{} ranked highest were resolved by the planner far more often than its lowest-ranked (98.75\% versus 51.00\%) and yielded shorter, more balanced routes, so the surrogate reserves explicit planning for the most promising molecules. We release nearly 20,000 retrosynthetic routes and the trained ensemble to support further work into the computer-aided synthesis planning of large, complex molecules.
 
\method{} captures route depth and modular balance rather than experimental feasibility, and validation against executed syntheses remains the natural next step. Even so, it offers a fast and interpretable way to connect virtual screening with the route characteristics that make a PROTAC practical to build.

\subsection*{AI use statement}

Generative AI tools were used to edit manuscript prose, propose organization and figure designs, identify potentially relevant literature, and provide critical feedback on statistical reporting and experimental design. They were not used to generate the reported experimental data or numerical model results. The authors reviewed the AI-assisted material, will verify every citation against the primary source, and take responsibility for the final text, claims, code and analyses.

\subsection*{Ethics statement}

This study uses molecular structures and computationally generated route descriptors and does not involve human participants, personal data or animal experiments. Synthetic-accessibility predictions can be over-interpreted as guarantees of laboratory feasibility. We mitigate this risk by defining the target explicitly, reporting its planner dependence and residual calibration, and requiring retrosynthetic and expert review before experimental decisions. Data, software and model releases will follow the licences and access restrictions of their upstream sources.

\subsection*{Reproducibility statement}

The experimental pipeline uses fixed data assignments, five recorded partition seeds, containerized environments and cached molecular features. The appendix specifies the target, grouping, optimization, metrics and ensemble algorithms. An anonymous supplementary archive containing source code, environment/container recipes, configuration files, split assignments and commands for dataset assembly, models training and evaluation, and plotting can be found here: \url{https://github.com/ribesstefano/RetroTAC} .

\subsubsection*{Author Contributions}

SR and RM conceptualized the idea, supervised and coordinated the work, and wrote the manuscript.
SR trained and evaluated the final models.
JAZ and TM conceptualized the synthesizability score, prototyped the models and training pipeline, and worked on the software.
LH tested the software.
KP and AP ran ShallowTree, and provided valuable discussions and insights on the synthesizability score and training pipeline.

\subsubsection*{Acknowledgments}

SR acknowledges funding provided by the Chalmers Gender Initiative for Excellence (Genie), and RM acknowledges funding provided by the Wallenberg AI, Autonomous Systems, and Software Program (WASP), supported by the Knut and Alice Wallenberg Foundation.
Computational resources were provided by the National Academic Infrastructure for Supercomputing in Sweden (NAISS), funded by the Swedish Research Council. Computational resources were also provided on the Berzelius system funded by the Knut and Alice Wallenberg foundation and operated by NAISS.

\bibliographystyle{unsrtnat}
\bibliography{references}

\newpage
\appendix

\section{Data distributions}

\subsection{Held-out clustering}
\label{app:heldout_clustering}

\begin{figure}[h]
    \centering
    \includegraphics[width=0.75\linewidth]{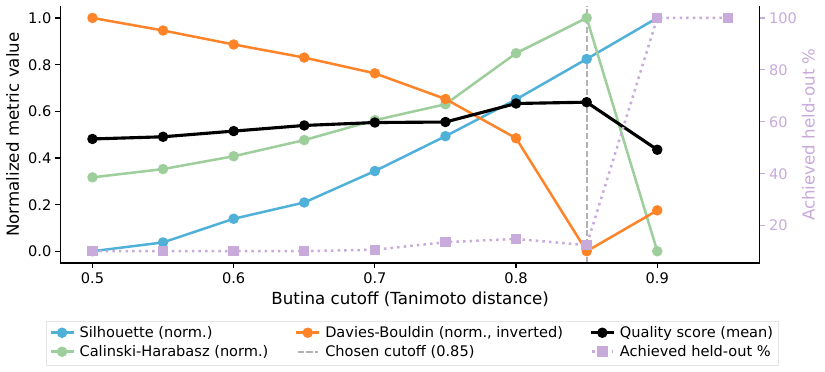}
    \caption{Cluster quality against Butina cutoff. Silhouette and Calinski--Harabasz are min-max normalized, Davies--Bouldin normalized and inverted, and the black curve is their mean. The right axis gives the attainable held-out fraction. The dashed rule marks the adopted cutoff of 0.85.}
    \label{fig:butina}
\end{figure}

We evaluated Butina Tanimoto-distance cutoffs from 0.50 to 0.95 in increments of 0.05 using 2,048-bit, radius-8 Morgan fingerprints of all 18,076 scored PROTACs. From 0.50 to 0.85, silhouette and Calinski--Harabasz scores improved, whereas the Davies--Bouldin index increased, indicating poorer quality by that criterion. We combined the three indices by min--max normalizing each, inverting Davies--Bouldin so that higher values consistently indicate better quality, and taking their mean (Figure~\ref{fig:butina}).

The cutoff also affected the held-out fraction because selection operated on complete clusters. The selection procedure returned approximately 10\% of molecules at cutoffs up to 0.70, 12--15\% at 0.75--0.85, and 100\% at 0.90 and 0.95. At 0.90, a single cluster contained 18,066 of the 18,076 molecules. We therefore restricted selection to cutoffs yielding a held-out fraction within three percentage points of the 10\% target and chose the highest composite quality among them. The selected cutoff, 0.85, produced 49 clusters in total and a held-out set of 2,232 molecules (12.3\%).

\subsection{Cross-validation fold distributions}
\label{app:fold_distributions}

\begin{figure}[t]
    \centering
    \includegraphics[width=\linewidth]{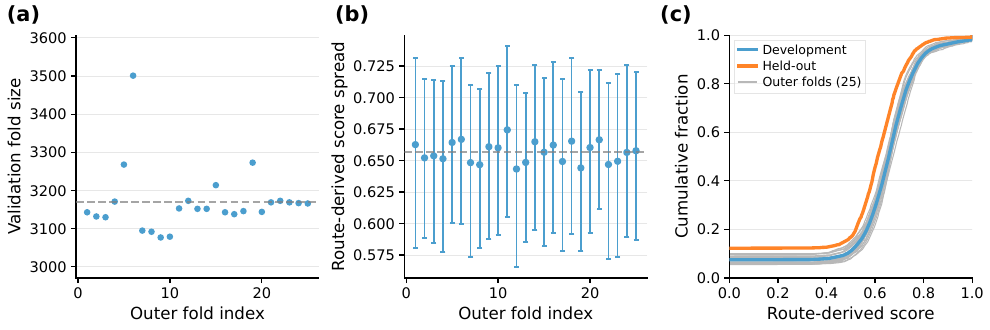}
    \caption{Distributions across the 25 outer validation folds (five seeds $\times$ five folds). (a) Fold sizes, with an equal split ($n/5\approx3,169$) shown by the dashed line. (b) Fold medians and interquartile ranges of the route-derived score; the dashed line marks the development-set median. (c) Empirical cumulative distribution functions for the validation folds (grey), development set (blue) and held-out set (orange).}
    \label{fig:folds}
\end{figure}

We reconstructed the 25 outer validation folds using the grouping function and seeds used for training. The 15,844 development molecules comprised 2,327 warhead scaffold groups, the largest containing 609 molecules (3.8\%). Despite this variation in group size, greedy assignment produced folds of similar size: 3,169$\pm$46 molecules and 465$\pm$62 groups, with every fold within 5\% of an equal split (Figure~\ref{fig:folds}a).

Label distributions were also consistent across folds. All fold medians were within 0.03 of the development-set median, which lay within every fold's interquartile range (Figure~\ref{fig:folds}b). Fold means ranged from 0.596 to 0.648 around the development mean of 0.621, while the proportions at $y=0$ and $y=1$ were 7.4$\pm$1.3\% and 1.9$\pm$0.4\%, respectively. The empirical cumulative distribution functions closely followed that of the development set (Figure~\ref{fig:folds}c), with a mean Kolmogorov--Smirnov statistic of 0.054. The held-out distribution differed more substantially (0.181), including a larger proportion of zero scores.

\section{Component leakage and cross-validation grouping}
\label{appendix:scaffold}

We evaluated twelve grouping keys on the 15,844 development-set PROTACs: the whole molecule, each component, and their pairwise and three-way combinations. We used generic Bemis--Murcko molecular scaffolds as grouping keys \citep{bemis_properties_1996}, with exact-SMILES fallback for acyclic inputs. Composite keys grouped molecules either by joint component identity or by connected components linking molecules that shared any constituent key.
Component overlap was measured across the same five seeds and five outer folds used during training.
No identical full-PROTAC SMILES occurred in both training and validation under any grouping strategy; the overlap reported here concerns individual components.

Whole-molecule grouping separated complete molecular representations but retained substantial component overlap (Table~\ref{tab:leakage}): 79.2\% of validation warheads, 71.8\% of linkers, and 89.2\% of E3 ligands also occurred in training as exact SMILES matches.
When two or three component types were required to be disjoint across folds, molecules linked directly or indirectly through any shared component key had to remain in the same group. The largest resulting group contained 93.2--99.9\% of the dataset, precluding approximately balanced five-fold CV.
Although whole-molecule groups had a higher association with the route-derived score than warhead groups ($\eta^2=0.884$ versus 0.575), they formed a much finer partition: 11,686 groups, with 59.2\% of molecules in singleton groups, compared with 3,128 groups and 10.6\% under warhead grouping.
Here, $\eta^2$ measures the fraction of observed score variance attributable to differences between group means; its dependence on partition granularity prevents interpreting a higher value as stronger protection against component leakage.

We therefore selected warhead grouping: it eliminated warhead overlap, retained manageable group sizes (largest group: 434 molecules, or 2.7\%), and had the highest observed association with the route-derived score among individual components. 
This grouping also evaluates generalization to unseen warhead groups while allowing linkers and E3 ligands already encountered during training.

\begin{table*}[t]
\centering
\small
\caption{Component overlap for representative grouping keys, reported as mean $\pm$ s.d.\ (\%) across 25 outer CV splits. Scaffold overlap uses generic Bemis--Murcko molecular scaffolds; exact-SMILES overlap measures matching component strings. Exact full-PROTAC SMILES overlap was zero under every grouping strategy. ``All components'' denotes connected-component grouping across warhead, linker, and E3 ligand keys; its largest group contains 99.9\% of the dataset. The adopted grouping key is shown in bold.}
\label{tab:leakage}
\begin{tabular}{lccc|ccc}
\toprule
& \multicolumn{3}{c}{Scaffold overlap (\%)} & \multicolumn{3}{c}{Exact-SMILES overlap (\%)} \\
\cmidrule(lr){2-4}\cmidrule(lr){5-7}
Grouped by & Warhead & Linker & E3 ligand & Warhead & Linker & E3 ligand \\
\midrule
Whole molecule & $84.6\pm0.6$ & $83.8\pm0.8$ & $93.4\pm0.6$ & $79.2\pm0.7$ & $71.8\pm0.8$ & $89.2\pm0.9$ \\
\textbf{Warhead} & $0.0\pm0.0$ & $77.9\pm3.3$ & $91.6\pm2.6$ & $0.0\pm0.0$ & $64.0\pm3.7$ & $86.8\pm2.9$ \\
Linker & $81.4\pm2.8$ & $0.0\pm0.0$ & $92.4\pm3.0$ & $75.7\pm3.1$ & $0.0\pm0.0$ & $87.7\pm4.1$ \\
E3 ligand & $76.5\pm5.1$ & $79.0\pm5.3$ & $0.0\pm0.0$ & $70.2\pm5.1$ & $64.1\pm6.2$ & $0.0\pm0.0$ \\
All components & $0.0\pm0.0$ & $0.0\pm0.0$ & $0.0\pm0.0$ & $0.0\pm0.0$ & $0.0\pm0.0$ & $0.0\pm0.0$ \\
\bottomrule
\end{tabular}%
\end{table*}

\section{Score design and analysis}
\label{app:score_analysis}

We examined the choice of score components and the sensitivity of the adopted two-term route-derived score to its weights, saturation depth and length-decay function. Of the 18,076 scored PROTACs, 16,318 had scorable routes; the remaining 304 purchasable and 1,454 unsolved targets retained fixed scores of 1 and 0 under every variant. The analyses below exclude these fixed endpoints. Including them increased correlations by 0.02--0.11 without changing the qualitative conclusions.

\begin{figure}[t]
    \centering
    \includegraphics[width=\linewidth]{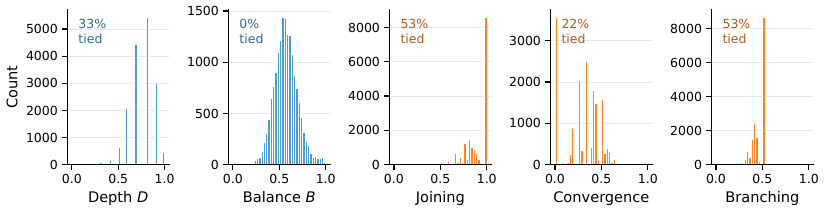}
    \caption{Distributions of the five candidate score components across 16,318 scorable routes. Blue denotes the adopted components; orange denotes the alternatives. Tied is the percentage of routes sharing each component's most common value.}
    \label{fig:score-terms}
\end{figure}

\paragraph{Candidate components.}
Fragment balance provided the finest resolution, with 13,375 distinct values and no value shared by more than 0.14\% of routes. The depth term was coarser, with five effective levels (Table~\ref{tab:score-terms}, Figure~\ref{fig:score-terms}). Joining fraction and rescaled branching assigned their most common value to 52.7\% and 53.1\% of routes, respectively; branching also had approximately half the standard deviation of either adopted term. Convergence was negatively correlated with the adopted score ($\rho=-0.297$), consistent with deeper routes accumulating more off-path steps. These distributions support using depth and balance to distinguish routes in this dataset: joining is already common, whereas the relative sizes of the joined fragments vary substantially.

\begin{table}[t]
\centering
\caption{Candidate score components across 16,318 scorable routes. Tied is the percentage sharing the most common score; effective levels is the exponential of the Shannon entropy of the score distribution. $R^2$ from adopted measures how well a component is explained by regression on route depth and balance. The adopted components are bold.}
\label{tab:score-terms}
\small
\resizebox{\linewidth}{!}{%
\begin{tabular}{lrrrrrrr}
\toprule
Term & Mean & s.d. & IQR & Tied (\%) & Eff.\ levels & $R^2$ from adopted & $\rho$ with $y$ \\
\midrule
\textbf{Depth $D$}    & 0.752 & 0.124 & 0.100 & 33.4 & 5.0 & --- & \phantom{$-$}0.723 \\
\textbf{Balance $B$}   & 0.568 & 0.126 & 0.160 & \phantom{0}0.1 & 11,995.3 & --- & \phantom{$-$}0.755 \\
Joining fraction       & 0.896 & 0.124 & 0.200 & 52.7 & 7.1 & 0.260 & \phantom{$-$}0.305 \\
Convergence            & 0.291 & 0.185 & 0.229 & 21.9 & 11.9 & 0.194 & $-$0.297 \\
Branching (rescaled)   & 0.453 & 0.065 & 0.100 & 53.1 & 7.5 & 0.203 & \phantom{$-$}0.265 \\
\bottomrule
\end{tabular}%
}
\end{table}

\paragraph{Sensitivity to parameter choices.}
Rank agreement remained high when the two adopted components were retained, with Spearman correlations of 0.906--0.997 across the tested parameter settings (Table~\ref{tab:score-variants}). Changing the saturation depth from $D_\mathrm{max}=11$ to 9 or 13 had little effect on ranking ($\rho\geq0.995$); larger changes to 7 or 15 retained $\rho\geq0.974$. Unequal length/balance weights of 0.25/0.75 and 0.75/0.25 had a greater effect ($\rho=0.917$ and 0.906). Logarithmic length decay largely preserved ranking ($\rho=0.989$) but shifted the score distribution, with a mean absolute change of 0.122 and a maximum of 0.139. At the fixed classification threshold of 0.7, this shift changed 25.1\% of labels, showing that rank agreement does not imply stability of thresholded classifications.

\paragraph{Sensitivity to component selection.}
Changing the component set generally had a larger effect on ranking. Adding branching with equal weights retained the highest agreement ($\rho=0.944$), whereas adding joining at weights of 0.50/0.30/0.20 reduced agreement to 0.791; including all five components gave 0.595 (Table~\ref{tab:score-variants}). Component selection therefore remains a substantive design choice: the adopted score prioritizes route depth and fragment-size balance, while the alternatives can substantially alter rankings.

\begin{table}[t]
\centering
\caption{Agreement between score variants and the adopted definition across 16,318 scorable routes. Label flips is the percentage whose binary label changes at a threshold of 0.7. Top-10\% $J$ is the Jaccard overlap between top-decile selections. Two-term weights are listed as length/balance; the three-term weights are length/balance/joining.}
\label{tab:score-variants}
\small
\begin{tabular}{lrrrrr}
\toprule
Variant & Spearman $\rho$ & Kendall $\tau$ & Mean $|\Delta|$ & Label flips (\%) & Top-10\% $J$ \\
\midrule
\multicolumn{6}{l}{\emph{Same two terms, re-parameterized}} \\
\midrule
Saturation $D_\mathrm{max}$ = 7    & 0.974 & 0.878 & 0.082 & 16.6 & 0.788 \\
Saturation $D_\mathrm{max}$ = 9    & 0.995 & 0.945 & 0.031 & \phantom{0}8.4 & 0.908 \\
Saturation $D_\mathrm{max}$ = 13   & 0.997 & 0.954 & 0.021 & \phantom{0}7.5 & 0.921 \\
Saturation $D_\mathrm{max}$ = 15   & 0.989 & 0.917 & 0.035 & 14.5 & 0.851 \\
Logarithmic depth decay           & 0.989 & 0.932 & 0.122 & 25.1 & 0.784 \\
Weights 0.25/0.75                & 0.917 & 0.762 & 0.050 & 13.4 & 0.657 \\
Weights 0.75/0.25                & 0.906 & 0.779 & 0.050 & 23.0 & 0.626 \\
\midrule
\multicolumn{6}{l}{\emph{Different term set}} \\
\midrule
Depth only                        & 0.723 & 0.588 & 0.100 & 50.1 & 0.289 \\
Balance only                       & 0.755 & 0.579 & 0.100 & 19.2 & 0.496 \\
All five terms, equal weights      & 0.595 & 0.435 & 0.080 & 31.1 & 0.259 \\
$+$ branching, equal weights       & 0.944 & 0.822 & 0.070 & 25.9 & 0.764 \\
$+$ joining, equal weights         & 0.853 & 0.716 & 0.082 & 37.9 & 0.695 \\
$+$ joining, 0.50/0.30/0.20      & 0.791 & 0.627 & 0.101 & 43.5 & 0.562 \\
$+$ convergence, equal weights     & 0.576 & 0.419 & 0.125 & 31.6 & 0.255 \\
\bottomrule
\end{tabular}
\end{table}

\paragraph{Implementation details.}
Fragment balance is averaged over bimolecular steps. Routes without such steps receive $B=0.5$, so their scores vary only with depth. This fallback is uncommon in the present dataset but could be more frequent in datasets with fewer fragment-joining reactions.

\subsection{Models and hyperparameter optimization}
\label{app:models}

\textbf{XGB.} XGBoost uses 1,024-bit Morgan fingerprints (radius 8, chirality enabled) concatenated with 2D RDKit descriptors. Hyperparameter optimization determines whether to apply truncated SVD to the fingerprints and the number of retained components. Descriptors are median-imputed, constant columns are removed, and the remaining columns are standardized. The target is quantile-transformed, and training uses the pseudo-Huber loss. All preprocessing is fitted only on the relevant training partition.

\textbf{MLP.} The PyTorch Lightning multilayer perceptron uses the same representation and preprocessing as XGBoost, with smooth L1 loss, a batch size of 256, a maximum of 500 epochs and early-stopping patience of 20 epochs.

\textbf{GNN.} The Chemprop directed message-passing network is initialized from CheMeleon, which was pretrained to predict Mordred descriptors from approximately one million PubChem structures \citep{graff2026chemprop,burns_deep_2025}. It receives molecular graphs parsed from SMILES and is fine-tuned end-to-end using mean squared error.

For each outer fold, Optuna's TPE sampler evaluates up to 30 XGB, 25 MLP and 15 GNN trials over the search space in Table~\ref{tab:hpo-space} \citep{akiba_optuna_2019}. Sampler seeds depend on the outer seed and fold index. A median pruner uses five startup trials and three warm-up steps; only the MLP reports intermediate objectives and is therefore eligible for pruning. Hyperparameters are selected on a grouped 80/20 split of the outer-training set by minimising
\begin{equation}
\mathcal{L}=0.5\,\mathrm{RMSE}+0.5\frac{1-\rho_s}{2},
\end{equation}
where $\rho_s$ is Spearman correlation. The selected configuration is refitted on the outer-training set and evaluated once on its validation fold. Final models are fitted on the full development set; the held-out set is excluded from hyperparameter selection.

\begin{table}[t]
\centering
\caption{Hyperparameter search space for the reported experiments. Log denotes logarithmic sampling; zero SVD components disables dimensionality reduction.}
\label{tab:hpo-space}
\small
\begin{tabular}{llll}
\toprule
Backend & Parameter & Type & Range \\
\midrule
XGB & \texttt{n\_estimators}      & int   & [200, 3000] \\
    & \texttt{max\_depth}         & int   & [3, 8] \\
    & \texttt{learning\_rate}     & float, log & $[10^{-3},\ 0.3]$ \\
    & \texttt{subsample}          & float & [0.6, 1.0] \\
    & \texttt{colsample\_bytree}  & float & [0.6, 1.0] \\
    & \texttt{min\_child\_weight} & int   & [1, 10] \\
    & \texttt{gamma}              & float & [0, 5] \\
    & \texttt{svd\_components}    & int   & [0, 128] \\
\midrule
MLP & \texttt{n\_layers}          & int   & [1, 4] \\
    & \texttt{hidden\_dim}        & categorical & \{64, 128, 256, 512\} \\
    & \texttt{dropout}            & float & [0, 0.5] \\
    & \texttt{learning\_rate}     & float, log & $[10^{-4},\ 10^{-2}]$ \\
    & \texttt{weight\_decay}      & float, log & $[10^{-6},\ 10^{-2}]$ \\
    & \texttt{svd\_components}    & int   & [0, 128] \\
\midrule
GNN & \texttt{ffn\_hidden\_dim}   & categorical & \{128, 256, 512\} \\
    & \texttt{ffn\_n\_layers}     & int   & [1, 3] \\
    & \texttt{dropout}            & float & [0, 0.4] \\
    & \texttt{max\_lr}            & float, log & $[10^{-4},\ 5\times10^{-3}]$ \\
\bottomrule
\end{tabular}
\end{table}

\begin{figure}[t]
    \centering
    \includegraphics[width=\linewidth]{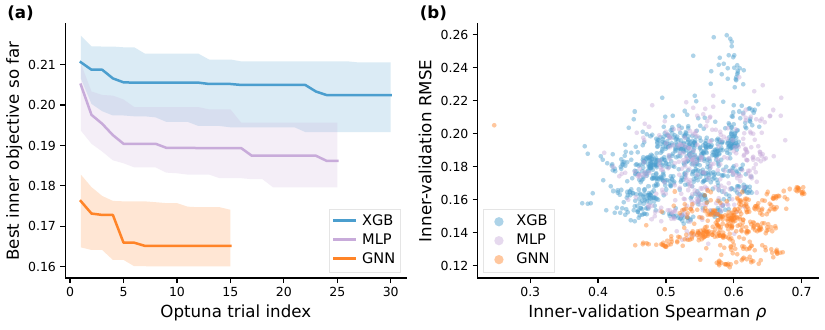}
    \caption{Hyperparameter search progress. (a) Best inner-validation objective by trial number, shown as the median and interquartile range across 25 outer folds. Traces end at each model's trial budget; pruned trials count towards the budget but cannot update the best objective. (b) Inner-validation RMSE and Spearman correlation for all completed trials.}
    \label{fig:hpo}
\end{figure}

Across the 25 outer folds, the search comprised 750 XGB, 625 MLP and 375 GNN trials; 52.5\% of MLP trials were pruned. By trial ten, all three models were within 2\% of their final best objective, with no subsequent improvement for the GNN in the reported trajectory (Figure~\ref{fig:hpo}a). Final objectives were 0.167$\pm$0.010 for GNN, 0.188$\pm$0.011 for MLP and 0.201$\pm$0.012 for XGB. The GNN achieved both lower RMSE and higher Spearman correlation during tuning (Figure~\ref{fig:hpo}b).

Selected hyperparameters varied across folds (Table~\ref{tab:hpo-selected}). For example, the XGB SVD dimension had an interquartile range of 41--92, and the most common MLP and GNN hidden dimensions were selected in fewer than half the folds. We therefore report their distributions alongside the final-model configurations rather than summarizing each search by a single setting.

\begin{table}[t]
\centering
\caption{Selected hyperparameters across 25 outer folds and the final-model configurations. Continuous parameters and larger integer ranges are summarized as median [interquartile range]; categorical parameters and small integer ranges are reported as mode (frequency).}
\label{tab:hpo-selected}
\small
\begin{tabular}{llrr}
\toprule
Backend & Parameter & Across the 25 folds & Final model \\
\midrule
XGB & \texttt{n\_estimators}      & 1,762 [1,280, 2,279] & 2,037 \\
    & \texttt{max\_depth}         & 7 (9/25)                   & 7 \\
    & \texttt{learning\_rate}     & $1.2\times10^{-2}$ [$5.0\times10^{-3}$, $1.9\times10^{-2}$] & $4.7\times10^{-3}$ \\
    & \texttt{subsample}          & 0.859 [0.718, 0.907]       & 0.822 \\
    & \texttt{colsample\_bytree}  & 0.727 [0.658, 0.858]       & 0.784 \\
    & \texttt{min\_child\_weight} & 4 (5/25)                   & 9 \\
    & \texttt{gamma}              & 2.30 [0.80, 3.14]          & 2.30 \\
    & \texttt{svd\_components}    & 62 [41, 92]                & 116 \\
\midrule
MLP & \texttt{n\_layers}          & 4 (16/25)                  & 4 \\
    & \texttt{hidden\_dim}        & 256 (8/25)                 & 128 \\
    & \texttt{dropout}            & 0.305 [0.214, 0.412]       & 0.189 \\
    & \texttt{learning\_rate}     & $2.6\times10^{-3}$ [$1.2\times10^{-3}$, $4.3\times10^{-3}$] & $5.0\times10^{-3}$ \\
    & \texttt{weight\_decay}      & $2.6\times10^{-4}$ [$3.2\times10^{-5}$, $1.4\times10^{-3}$] & $7.9\times10^{-5}$ \\
    & \texttt{svd\_components}    & 68 [36, 105]               & 36 \\
\midrule
GNN & \texttt{ffn\_hidden\_dim}   & 512 (10/25)                & 256 \\
    & \texttt{ffn\_n\_layers}     & 2 (11/25)                  & 1 \\
    & \texttt{dropout}            & 0.223 [0.124, 0.305]       & 0.298 \\
    & \texttt{max\_lr}            & $6.7\times10^{-4}$ [$3.3\times10^{-4}$, $1.4\times10^{-3}$] & $6.6\times10^{-4}$ \\
\bottomrule
\end{tabular}%
\end{table}

\section{Additional performance results}
\label{appendix:performance}

\subsection{Normality and variance diagnostics}
\label{app:normality}

Following the Polaris method-comparison protocol \citep{ash_practically_2025}, we assessed the suitability of parametric comparisons using variance diagnostics and the shape of the fold-score distributions. The protocol motivates ANOVA and Tukey HSD by their robustness to moderate departures from normality in balanced comparisons, with each model contributing 25 evaluations from $5\times5$ repeated cross-validation. Accordingly, we examined Q--Q plots and reported Shapiro--Wilk tests as supplementary diagnostics, rather than using normality-test rejection alone to select the testing procedure.

For variance homogeneity, we adopted the protocol's maximum-to-minimum variance-ratio threshold of 9 for equally sized groups \citep{blanca_effect_2018}. Our decision rule in Section \ref{sec:models} flags substantial heteroscedasticity only when this threshold is exceeded and Levene's $p<0.05$. None of the four metrics met this criterion (Table~\ref{tab:variance}): variance ratios ranged from 1.10 to 4.68 and Levene's $p$-values from 0.062 to 0.980. For $R^2$, the variance ratio was 4.6849 and Levene's $p=0.0877$.

\begin{figure}[t]
    \centering
    \includegraphics[width=\linewidth]{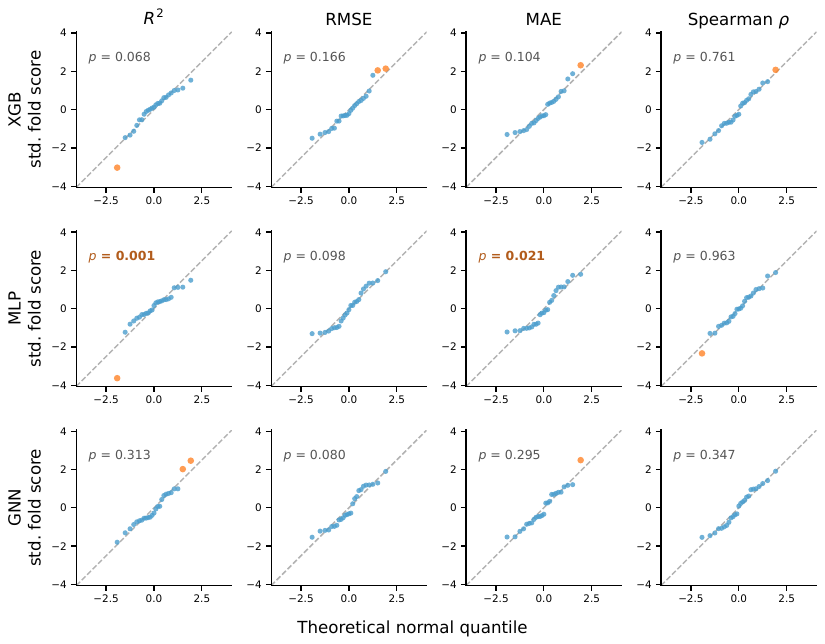}
    \caption{Normal Q--Q plots of the 25 outer-fold scores for each model and metric. Scores are standardized within panels, which share a common scale and the reference line $y=x$. Orange points lie more than two standard deviations from the mean. Annotations give Shapiro--Wilk $p$-values, with rejections at $\alpha=0.05$ in bold.}
    \label{fig:qq}
\end{figure}

Shapiro--Wilk tests rejected normality at $\alpha=0.05$ for two of the twelve model--metric combinations, both involving the MLP (Table~\ref{tab:normality}, Figure~\ref{fig:qq}). The $R^2$ rejection ($p=0.001$) was associated with one unusually low result for one particular fold, $R^2=-0.12$, approximately 3.6 standard deviations below the MLP mean. Excluding this observation for diagnostic purposes increased the Shapiro--Wilk $p$-value to 0.87. The MAE distribution also departed from normality ($p=0.021$), with negative excess kurtosis; removing its largest value did not eliminate the rejection ($p=0.026$). All 25 observations were retained in the statistical comparisons. We used repeated-measures ANOVA followed by Tukey HSD in accordance with the stated protocol, while documenting here these departures from normality.

\begin{table}[t]
\centering
\caption{Between-model variance diagnostics for 25 paired outer-fold evaluations. The stated decision rule selects the parametric procedure unless both Levene's $p<0.05$ and the maximum-to-minimum variance ratio exceeds 9.}
\label{tab:variance}
\small
\begin{tabular}{lrrrl}
\toprule
Metric & Levene $W$ & Levene $p$ & Max/min variance & Branch \\
\midrule
$R^2$           & 2.518 & 0.088 & 4.68 & parametric (forced) \\
RMSE            & 2.900 & 0.062 & 3.21 & parametric (forced) \\
MAE             & 1.702 & 0.190 & 2.39 & parametric (forced) \\
Spearman $\rho$ & 0.021 & 0.980 & 1.10 & parametric (forced) \\
\bottomrule
\end{tabular}
\end{table}

\begin{table}[t]
\centering
\caption{Shapiro--Wilk tests, sample skewness and excess kurtosis for the 25 outer-fold scores of each model. Rejections at $\alpha=0.05$ are bold.}
\label{tab:normality}
\small
\begin{tabular}{llrrrr}
\toprule
Model & Metric & Skew & Excess kurtosis & $W$ & $p$ \\
\midrule
XGB & $R^2$           & $-$1.204 & \phantom{$-$}2.182 & 0.925 & 0.068 \\
    & RMSE            & \phantom{$-$}0.674 & \phantom{$-$}0.004 & 0.942 & 0.166 \\
    & MAE             & \phantom{$-$}0.726 & $-$0.180 & 0.933 & 0.104 \\
    & Spearman $\rho$ & \phantom{$-$}0.200 & $-$0.735 & 0.975 & 0.761 \\
\midrule
MLP & $R^2$           & $-$1.922 & \phantom{$-$}6.521 & \textbf{0.842} & \textbf{0.001} \\
    & RMSE            & \phantom{$-$}0.302 & $-$1.153 & 0.932 & 0.098 \\
    & MAE             & \phantom{$-$}0.417 & $-$1.240 & \textbf{0.903} & \textbf{0.021} \\
    & Spearman $\rho$ & $-$0.143 & $-$0.053 & 0.985 & 0.963 \\
\midrule
GNN & $R^2$           & \phantom{$-$}0.682 & \phantom{$-$}0.472 & 0.954 & 0.313 \\
    & RMSE            & \phantom{$-$}0.252 & $-$1.286 & 0.928 & 0.080 \\
    & MAE             & \phantom{$-$}0.446 & $-$0.118 & 0.953 & 0.295 \\
    & Spearman $\rho$ & \phantom{$-$}0.111 & $-$1.145 & 0.956 & 0.347 \\
\bottomrule
\end{tabular}
\end{table}

\subsection{Additional regression and classification metrics}
\label{app:remaining-metrics}

\begin{figure}[t]
    \centering
    \includegraphics[width=\linewidth]{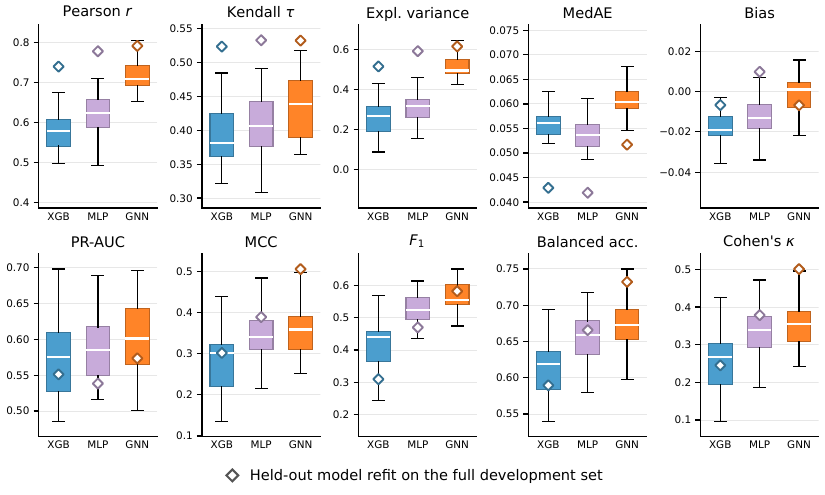}
    \caption{Additional performance metrics across cross-validation and held-out evaluation. Boxes summarize the 25 outer folds; open diamonds denote models refitted on the full development set and evaluated on 2,232 held-out molecules. The top row shows regression and ranking metrics; the bottom row shows classification metrics at a threshold of 0.7, with PR-AUC calculated from continuous predictions.}
    \label{fig:remaining}
\end{figure}

Tables~\ref{tab:mae-thresholded} and~\ref{tab:remaining} report metrics that complement the main-text results; Figure~\ref{fig:remaining} shows their fold distributions. In cross-validation, mean Pearson and Kendall correlations, explained variance, PR-AUC, MCC, $F_1$, balanced accuracy and Cohen's $\kappa$ all followed the ordering GNN $>$ MLP $>$ XGB.

Several metrics highlighted different aspects of performance. Specificity was highest for XGB (0.908), followed by MLP (0.852) and GNN (0.813), consistent with XGB's higher precision and lower recall. Overall accuracy was similar for MLP and GNN (0.722 versus 0.720), whereas balanced accuracy favored GNN (0.670 versus 0.655). Median absolute error was lowest for MLP and highest for GNN, despite the GNN's lower mean absolute and squared errors, suggesting that its advantage was concentrated among larger errors. Mean signed error was small for all models and closest to zero for GNN.

On the held-out set, most correlation and error metrics improved relative to cross-validation. These comparisons reflect both refitting on the full development set and a different evaluation distribution, so the improvement cannot be attributed to training-set size alone. PR-AUC decreased for all models, and $F_1$ decreased for XGB and MLP. Predicted-positive rates fell from 16.6--30.0\% in cross-validation to 5.6--14.7\% on the held-out set, where approximately 82\% of molecules were negative at the 0.7 threshold. The GNN nevertheless retained the highest held-out recall (0.53), MCC (0.505) and balanced accuracy (0.732).

\begin{table}[t]
\centering
\caption{MAE, precision and recall. CV values are mean $\pm$ s.d.\ over 25 paired outer evaluations; held-out values are evaluated on 2,232 molecules. Precision and recall use a threshold of 0.7 for both reference scores and predictions. The best value within each evaluation setting is bold.}
\label{tab:mae-thresholded}
\footnotesize
\begin{tabular}{llrrr}
\toprule
Evaluation & Model & MAE & Precision & Recall\\
\midrule
5$\times$5 CV & XGB & 0.104$\pm$0.010 & \textbf{0.63$\pm$0.06} & 0.31$\pm$0.10\\
& MLP & 0.099$\pm$0.009 & 0.61$\pm$0.04 & 0.46$\pm$0.08\\
& GNN & \textbf{0.094$\pm$0.007} & 0.59$\pm$0.04 & \textbf{0.53$\pm$0.12}\\
\midrule
Held-out & XGB & 0.085 & \textbf{0.66} & 0.20\\
& MLP & \textbf{0.076} & 0.58 & 0.39\\
& GNN & 0.086 & 0.65 & \textbf{0.53}\\
\bottomrule
\end{tabular}
\end{table}

\begin{table}[t]
\centering
\caption{Additional regression, ranking and classification metrics. CV values are mean $\pm$ s.d.\ over 25 paired outer evaluations. Held-out values are from models refitted on the full development set and evaluated on 2,232 molecules. Classification metrics use a threshold of 0.7; PR-AUC uses continuous predictions against the thresholded reference scores. The best value within each evaluation setting is bold, except for bias and predicted-positive rate.}
\label{tab:remaining}
\footnotesize
\begin{tabular}{lrrrrrr}
\toprule
 & \multicolumn{3}{c}{5$\times$5 CV} & \multicolumn{3}{c}{Held-out} \\
\cmidrule(lr){2-4}\cmidrule(lr){5-7}
Metric & XGB & MLP & GNN & XGB & MLP & GNN \\
\midrule
MedAE & 0.056$\pm$0.003 & \textbf{0.054$\pm$0.003} & 0.061$\pm$0.004 & 0.043 & \textbf{0.042} & 0.052 \\
Max error & 0.928$\pm$0.114 & 0.979$\pm$0.069 & \textbf{0.819$\pm$0.123} & \textbf{0.704} & 0.819 & 0.704 \\
Expl. variance & 0.250$\pm$0.116 & 0.298$\pm$0.105 & \textbf{0.512$\pm$0.054} & 0.516 & 0.592 & \textbf{0.616} \\
Bias & -0.020$\pm$0.013 & -0.013$\pm$0.012 & -0.001$\pm$0.011 & $-$0.007 & 0.010 & $-$0.007 \\
Pearson $r$ & 0.577$\pm$0.056 & 0.619$\pm$0.053 & \textbf{0.717$\pm$0.037} & 0.740 & 0.778 & \textbf{0.792} \\
Kendall $\tau$ & 0.393$\pm$0.043 & 0.408$\pm$0.045 & \textbf{0.432$\pm$0.044} & 0.523 & \textbf{0.533} & 0.532 \\
PR-AUC & 0.574$\pm$0.051 & 0.588$\pm$0.045 & \textbf{0.602$\pm$0.050} & 0.551 & 0.538 & \textbf{0.574} \\
MCC & 0.280$\pm$0.075 & 0.338$\pm$0.064 & \textbf{0.353$\pm$0.066} & 0.301 & 0.389 & \textbf{0.505} \\
$F_1$ & 0.410$\pm$0.105 & 0.519$\pm$0.062 & \textbf{0.548$\pm$0.081} & 0.309 & 0.470 & \textbf{0.582} \\
Specificity & \textbf{0.908$\pm$0.032} & 0.852$\pm$0.041 & 0.813$\pm$0.064 & \textbf{0.977} & 0.937 & 0.937 \\
Balanced acc. & 0.611$\pm$0.041 & 0.655$\pm$0.034 & \textbf{0.670$\pm$0.040} & 0.590 & 0.666 & \textbf{0.732} \\
Accuracy & 0.710$\pm$0.036 & \textbf{0.722$\pm$0.036} & 0.720$\pm$0.033 & 0.836 & 0.838 & \textbf{0.863} \\
Cohen's $\kappa$ & 0.251$\pm$0.086 & 0.329$\pm$0.067 & \textbf{0.347$\pm$0.072} & 0.245 & 0.379 & \textbf{0.501} \\
Pred. pos. rate & 0.166$\pm$0.055 & 0.251$\pm$0.055 & 0.300$\pm$0.083 & 0.056 & 0.123 & 0.147 \\
\bottomrule
\end{tabular}
\end{table}

\section{Comparison with DeepPSA}
\label{app:deeppsa}

We compared DeepPSA's predicted probability of being ``easy to synthesize'' \citep{zhang_deeppsa_2025} with the route-derived score and predictions from \method{}'s Caruana ensemble. From the held-out set, we removed 51 molecules found in DeepPSA's training data and a further 378 used for ensemble selection, leaving 1,803 for evaluation. Pearson and Spearman correlations used continuous outputs. For Kendall's $\tau$, route scores and \method{} predictions were thresholded at 0.7 and DeepPSA probabilities at 0.5.

Continuous-output correlations with DeepPSA were higher for \method{} predictions ($r=0.314$, $\rho=0.492$) than for the route-derived scores ($r=0.239$, $\rho=0.372$; Table~\ref{tab:deeppsa}). The corresponding thresholded Kendall correlations were similar (0.249 and 0.254, respectively).

\begin{table}[t]
\centering
\caption{Agreement with DeepPSA on the 1,803 held-out molecules remaining after exclusions for training overlap and ensemble selection. Pearson and Spearman correlations use continuous outputs; Kendall's $\tau$ uses outputs thresholded at 0.7 for the route-derived score and \method{}, and 0.5 for DeepPSA.}
\label{tab:deeppsa}
\small
\begin{tabular}{lrrr}
\toprule
Comparison & Pearson ($r$) & Spearman ($\rho$) & Kendall ($\tau$) \\
\midrule
Route-derived score vs. DeepPSA & 0.2392 & 0.3724 & 0.2539 \\
Caruana prediction vs. DeepPSA & 0.3144 & 0.4919 & 0.2485 \\
\bottomrule
\end{tabular}
\end{table}

\section{Additional example routes}
\label{app:routes}

\begin{figure}[t]
    \centering
    \includegraphics[width=\linewidth]{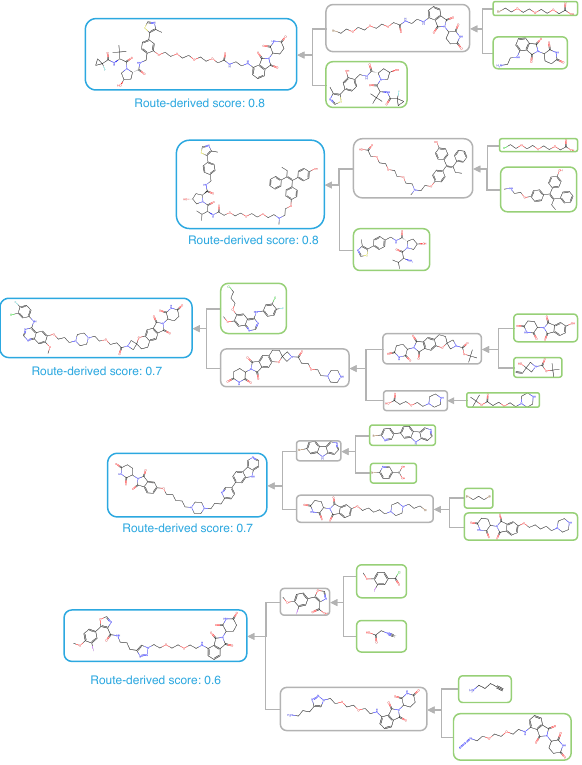}
    \caption{Five retrosynthesis routes generated by ShallowTree.}
    \label{fig:example_routes}
\end{figure}

Figure~\ref{fig:example_routes} shows five routes generated by ShallowTree, with route-derived synthesizability scores ranging from 0.6 to 0.8. Target PROTACs are outlined in blue, intermediates in grey, and purchasable building blocks in green. The examples include sequential fragment assembly and convergent routes that join separately prepared intermediates, illustrating how different route architectures can achieve similar scores through their combinations of route depth and fragment balance.

\end{document}